\documentclass[twocolumn]{article}
\usepackage{times}
\usepackage{fullpage}
\usepackage{xspace}
\usepackage{url}
\usepackage{graphicx}

\def\sql{{\sc sql}\xspace}
\def\nosql{{\sc nosql}\xspace}
\def\mysql{{\sc mysql}\xspace}
\def\mosql{{\sc mosql}\xspace}
\def\newsql{{\sc newsql}\xspace}
\def\eg{e.g.}

\begin{document}
\title{Yesquel: scalable SQL storage for Web applications}
\author{
  Marcos K. Aguilera$^{\ast}$ \quad
  Joshua B. Leners$^{\dag}$ \quad
  Ramakrishna Kotla$^{\ast}$ \quad
  Michael Walfish$^{\ddag}$ \\
  $^{\ast}$unaffiliated \quad\quad 
  $^{\dag}$UT Austin \quad\quad
  $^{\ddag}$NYU
}
\date{8 November 2014}

\maketitle
\begin{abstract}
Based on a brief history of the storage systems for Web applications,
we motivate the need for a new storage system.
We then describe the architecture of such a system, called Yesquel.
Yesquel supports the \sql query language and offers performance similar to
\nosql storage systems.
\end{abstract}

\section{Introduction}

Web applications (web mail, web stores, social networks, etc)
keep massive amounts of data
(account settings, user preferences, passwords,
emails, shopping carts, wall posts, etc). The design of such
an application often revolves around the underlying storage system.

This paper briefly describes a new
storage system for Web applications, called Yesquel.
Yesquel combines several advantages of prior systems: it
supports the \sql query language to facilitate the design of
applications, and it offers performance similar to \nosql
key-value storage systems.

To achieve these goals, Yesquel leverages techniques for
scalability and fault tolerance previously used in \nosql systems,
and uses them to obtain a \sql system.
In addition, Yesquel incorporates a unique
architecture that provides an embedded query processor to each of its
clients, that implements distributed transactions at a low level,
and that builds a distributed index structure on top of such
transactions. This architecture is accompanied by new and efficient
mechanisms to execute transactions and to store database
tables and indexes.

This paper focuses on Yesquel's motivation and architecture.
A subsequent paper will describe Yesquel's transactions and
storage engine.

\section{Historical perspective and\\motivation}

The storage systems used in Web applications have evolved
dramatically over the past 25 years, and the history
brings interesting insights. We can roughly divide
these storage systems into four generations. 

\begin{itemize}

\item {\em First generation: file systems.} In the early 1990s,
  Web pages were static. Web servers
  received HTTP requests for a file,
  read the file from their local file system, and returned it to the
  user. The storage system for Web applications
  was simply the file system holding such files.

\item {\em Second generation: file and\/ \sql database systems.} In the mid to late 1990s,
  the Web saw the emergence of {\em dynamic content}---content that depends on
  who the user is or what the user has done (\eg, the items in
  a shopping cart). To generate a page, the Web server invoked a program
  written in languages
  such as Perl, Python, PHP, Java, etc. The program stored
  data needed to generate the page in a central
  \sql database system.
  The use of \sql was convenient, because
  \sql has many useful features (joins, secondary indexes,
  transactions, aggregations, many data types, etc).
  The storage system,
  thus, was a combination of the file system for static content
  and the database system for dynamic content.

\item {\em Third generation: highly scalable systems.} In the 2000s,
  large Web sites emerged, such as Amazon,
  Hotmail, Google, Yahoo, and others. These sites
  had a rapid growth in the number of users; soon the
  database system became a performance and scalability
  bottleneck.
  Scaling the database system was hard or expensive, so developers
  decided to replace the \sql
  database system with their own home-grown storage systems,
  such as the Google File System~\cite{GGL2003}, BigTable~\cite{bigtable},
  Dynamo~\cite{dynamo2007}, and others. This was the beginning
  of a movement that later became known as \nosql.

\item {\em Fourth generation: cloud storage systems.} In the 2010s, many
  Web applications moved to the cloud, where
  many vendors share the same computing and storage
  infrastructure. Storage systems were designed not just to scale, but
  also to provide isolation, so that applications do
  not interfere with one another. Examples of such storage
  systems include Amazon S3, SimpleDB, Azure Blobs, and Azure Tables.
  
\end{itemize}

\sloppypar
A highlight in this history is the emergence of the
\nosql movement, ten years ago, which sought to replace the \sql database
system with cheaper custom-built alternatives.
These alternatives were much simpler than a \sql database system,
and thus were easier to scale, but they offered more restricted
functionality: few \nosql systems offer transactions,
secondary indexes, joins, or aggregations, and certainly no
\nosql systems offer all the features found in \sql.
Thus, the move from the \sql database system to \nosql systems
came at the cost of functionality.

Today, there are several dozens of \nosql systems, each offering
a different set of features, and each with its own application
interface. Web application developers face a difficult
choice of what storage system to use, because it is unclear a priori
what features the application might need. Even if developers can identify
an appropriate system, the application may evolve and later require
functionality from the storage system that was not originally deemed useful.
Because each storage system provides its own
interface, once an application is developed to use one
storage system, it becomes hard to replace it
with another storage system should the need arise---a problem known as
vendor lock-in.

We start with the observation that \nosql is not a feature
but rather the absence of a feature. What is the \nosql
movement really trying to achieve? The answer is a
distributed storage system that is scalable; fault tolerant;
simple and sane to design, implement, and maintain;
and nimble and cheap to run.

We took these characteristics as our goal in developing Yesquel, but
we also wanted to provide full support
for the \sql query language. Besides having a rich set of
features, \sql is a well-known query language (it is taught in
database courses), and it is an industry
standard, so applications developed for \sql avoid the problem of
vendor lock-in.

\section{Architecture}

A \sql database system has two main components:
  a query processor and a storage engine.
The query processor parses and executes \sql queries, while
  the storage engine stores tables and indexes.

\begin{figure}
  \centerline{
    \includegraphics{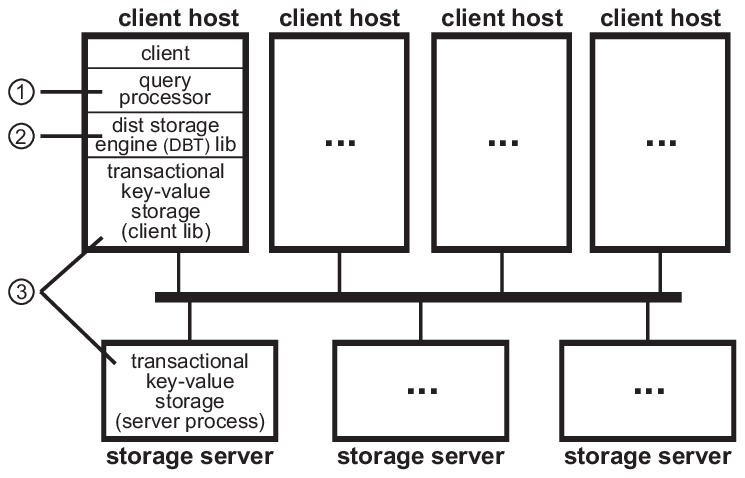}
  }
  \caption{Architecture of Yesquel.}
  \label{archfig}
\end{figure}

The architecture of Yesquel is depicted in Figure~\ref{archfig}.
Each client has its own embedded query processor
(box \textcircled{\small 1} in the figure), which is a
  user library that links against the client application.
As a result, as the number of clients increases,
  the number of query processors also increases proportionately.
This property allows Yesquel to scale the query processing
  capacity of the system.
The query processors all share a common storage engine.

For this idea to work, the storage engine must be designed
  to handle a large number of query processors, which
  at a given time may try to access the same tables and indexes.
Handling concurrent access to such data is a key challenge
  addressed by Yesquel.
  
In Yesquel, the storage engine is implemented as a
  {\em distributed balanced tree (DBT)} (box~\textcircled{\small 2}).
A DBT is a balanced tree data structure whose nodes are
  distributed over a set of storage servers.
The reason for distribution is to scale the performance
  of the DBT; in Yesquel, this is done by increasing
  the number of storage servers.

\sloppypar  
The Yesquel DBT library is implemented on top of a transactional
  key-value storage system---a simple transactional system that
  stores key-value pairs on the storage servers (boxes~\textcircled{\small 3}).
That is, the nodes of the Yesquel DBT are stored as
  key-value pairs in the key-value storage system.
This design separates the implementation of the DBT
  from the implementation of the distributed
  transactions.
The transactions in the key-value storage system provide
  {\em snapshot isolation}~\cite{si}, a property often used in
  commercial database systems.
Because the DBT is implemented above the transactions, the DBT
  benefits from the full power of transactions; for example,
  the Yesquel DBT uses transactions to atomically move
  data across DBT nodes to balance load.

Note that transactions are provided at the
  lowest layer---the key-value storage system, which is the layer that
  actually stores
  the data bits on storage servers.
As a result, the transactional protocol
  enjoys greater efficiency.
Specifically, Yesquel uses multi-version concurrency
  control---the most sophisticated form of
  concurrency control---which requires managing
  multiple versions of each data item, and this can be
  done most efficiently and effectively at the layer
  that stores the actual data.

\section{Related work}

F1~\cite{f1google2013} is a distributed storage system designed for Google's AdWords;
  it provides support for \sql
  with a non-relational hierarchical model.
F1 has a different architecture from Yesquel: F1
  is layered on top of the Spanner system, which
  in turn is layered onto a Bigtable-based implementation.
Spanner~\cite{spanner2012} provides distributed transactions, while
  Bigtable provides the DBT functionality.
Thus, the DBT is implemented {\em below} transactions.  
In contrast, in Yesquel distributed transactions
  are provided at the lowest level (the key-value storage system),
  and the DBT is implemented {\em above} the transactions.
As explained, this choice allows the DBT to leverage
  transactions.
Moreover, Yesquel provides an embedded query processor to clients.
Yesquel also uses different protocols for transactions, which
  unlike F1 do not require special hardware clocks.
However, F1 provides a stronger transaction isolation
  property (strict serializability) than Yesquel, and
  F1 works in a geo-distributed deployment, whereas Yesquel
  runs within a single data center.
  
\mosql~\cite{mosql2011,mosql2013} is a distributed storage engine for \mysql.
\mosql has a different architecture from Yesquel: its transactional
  layer runs on top of (1) a distributed storage layer without transactions,
  and (2) a certifier service implemented as a replicated state
  machine.
Atop the transaction layer, \mosql has a B+tree.
In contrast, Yesquel provides distributed
  transactions at the lowest level, Yesquel uses a more
  sophisticated DBT, and Yesquel does not use a logically
  centralized certifier.
As a result, we believe Yesquel is more efficient and scalable than \mosql.
However, \mosql provides a stronger transaction
  isolation (serializability) than Yesquel.

Traditionally, there are two broad architectures for a
    distributed \sql database system:
    shared-nothing and shared-disk~\cite{LM1975}.
Each has advantages, and there is a long-running debate
    about the two approaches, with commercial vendors
    supporting one or another, and sometimes both.

\sloppypar
In \emph{shared-nothing} systems
  (Clustrix,
  Greenplum,
  H-Store/VoltDB~\cite{hstore},
  IBM DB2 DPF,
  Microsoft SQL Server,
  MySQL Cluster,
  Netezza,
  Tandem NonStop, 
  Teradata, etc.),
    each database server stores part of the data; the system decomposes 
    queries, executes the sub-queries at the appropriate servers, and
    combines the results for the client.
The benefits of this architecture can be substantial;
    for example, sub-query processing happens close to the data,
    thereby avoiding network communication.
However, performance crucially depends on a good partition of the data,
    and such a good partition may not exist (if the query set is
    dynamic).
Or a partition may be expensive to 
    identify~\cite{caseforsharednothing1986}:
    the classical
    approach is to rely on a (well-paid)
    database administrator to partition manually.
Software can help (as in H-Store/VoltDB~\cite{hstore,skewaware})
   but not in all cases.
    
In \emph{shared-disk} systems
  (IBM DB2 pureScale,
  Oracle RAC,
  ScaleDB, 
  Sybase ASE Cluster Edition, etc.),
every database server can access every data item; the disks
    are shared over the network.
The advantage is easier management as there is no need to carefully
  partition the database.
However, in this architecture, servers must carefully coordinate
    access to storage; the traditional solutions use
  distributed locks, distributed leases,
  and cache coherence protocols, which bring complexity and cost.

A recent movement called \newsql advocates new architectures
  for database systems.
\newsql systems include H-Store/VoltDB and Hekaton.
H-Store/VoltDB is an in-memory shared-nothing system, where each server
  runs a single thread to avoid synchronization overheads.
Similar to other shared-nothing systems, performance critically
  depends on a good partition of the data.
Hekaton is a centralized in-memory database system that features
  a lock-free index structure; it performs
  well, but scalability is limited to a single machine.

Finally, there has been much work on Big Data systems that
   process massive data sets at many hosts
   (\eg, \cite{hbase,hadoop,mapreduce}).
Some of these systems support \sql~\cite{shark,asterdata}.
However, they are intended for data analytics applications and
    are not directly applicable to Web applications.

\section{Summary}

We described the architecture of
  Yesquel, a storage system designed for Web applications,
  whose workload consists of relatively small queries that must execute
  quickly.
Yesquel provides full support for \sql
  and offers performance similar to \nosql key-value storage systems.
Basically, Yesquel maps \sql queries to operations on a DBT, which
  in turn are mapped to operations on a transactional key-value
  storage system.
In other words, Yesquel internally leverages a \nosql storage
  system to provide its scalability.

\bibliography{bib}
\bibliographystyle{abbrv}
\end{document}